\documentclass[journal]{IEEEtran}
\usepackage{amsmath}
\usepackage{amsfonts}
\usepackage{amstext}
\usepackage{epsfig}
\usepackage{amssymb}
\usepackage{cite}
\usepackage{hhline}
\usepackage{multirow}
\usepackage{framed}
\usepackage{xcolor}
\usepackage{lipsum}
\usepackage{enumerate}
\usepackage{color}
\usepackage{caption}
\usepackage{subcaption}
\usepackage{graphicx}
\usepackage{algorithm}
\usepackage{algpseudocode}
\usepackage{mathtools}
\usepackage{booktabs}
\usepackage{bm}
\usepackage{cite}
\usepackage{url}
\usepackage{makecell}

\usepackage[utf8]{inputenc}
\usepackage[english]{babel}

\usepackage{breqn}
\usepackage{makecell}

\begin{document}
\title{	Efficient Beamforming for Mobile mmWave Networks}
\author{\IEEEauthorblockN{Sara Khosravi, Hossein S. Ghadikolaei, and Marina Petrova}\\
\IEEEauthorblockA{School of EECS, KTH Royal Institute of Technology, Stockholm, Sweden \\
	Emails:\{sarakhos, hshokri, petrovam\}@kth.se}
}

\maketitle

\newcommand{\RNum}[1]{\lowercase\expandafter{\romannumeral #1\relax}}
\begin{abstract}

We design a lightweight beam-searching algorithm for mobile millimeter-wave systems. We construct and maintain a set of \textit{path skeleton}s, i.e., potential paths between a user and the serving base station to substantially expedite the beam-searching process. To exploit the spatial correlations of the channels, we propose an efficient algorithm that measures the similarity of the skeletons and re-executes the beam-searching procedure only when the old one becomes obsolete. We identify and optimize several tradeoffs between: \mbox{\RNum{1}) the} beam-searching overhead and the instantaneous rate of the users, and \RNum{2}) the number of users and the update overhead of the path skeletons. Simulation results in an outdoor environment with real building map data show that the proposed method can significantly improve the performance of beam-searching in terms of latency, energy consumption and achievable throughout.
\end{abstract}


\section{Introduction}
\label{In}
Millimeter-wave (mmWave) communication is regarded as a promising solution to support high data rate demands in the next generations of wireless networks \cite{rappaport2013millimeter}. 
To compensate for the severe propagation loss in this band, both transmitters (Txs) and receivers (Rxs) rely on directional communications using a relatively large number of antennas, feasible thanks to the short wavelength. The use of directional communications, especially with analog or hybrid beamforming architectures \cite{kutty2016beamforming}, complicates the channel estimation and beamforming tasks, since the channel dimension is large and it is available through the eyes of the analog filters (which are low-rank and non-invertible). The problem becomes even more challenging in a mobile network, where the mobility demands frequent re-executions of the optimal beamforming task to overcome misalignment between Tx and Rx beams \cite{shokri2015millimeter}. Therefore, there is a natural tradeoff between the total beamforming overhead (which is a function of the number of re-executions of the task) and the instantaneous rate of the user. 

To address the problem in a stationary environment, the existing approaches usually search over a codebook (a set of potential beams) to find the optimal beam pairs for Tx and Rx. In particular, the  
existing mmWave standards \cite{5174147,6171799} define a multi-resolution codebook and use an exhaustive beam-searching algorithm to find the direction with the maximum link budget. However, this approach is time-consuming because it includes many iterations of sending pilot signals and waiting for control/feedback frames. Generally speaking, the overhead  increases with the number of beam directions and Tx/Rx may remains most of the time in the beam scanning phase rather than the data transmission phase \cite{sur2016beamspy}.
Other approaches, such as sparsity-aware beamforming \cite{heath2016overview} or subspace estimation \cite{ghauch2016subspace}, face a similar problem: their overheads hinder their applicability in mobile mmWave networks. 
Although the recent compressive-sensing based approaches \cite{marzi2016compressive,heath2016overview,Hassanieh} need a logarithmic number of measurements in beam-searching, they \mbox{do not} work with the existing mmWave devices because they require an adopted phase-array antennas \cite{Hassanieh} or phase coherent measurements \cite{marzi2016compressive}.

The spatial channel response (SCR) of mmWave systems has two fundamental properties. First, it is sparse in the angular domain  \cite{sur2016beamspy,rappaport2013broadband,heath2016overview,8057188}. In other words, there are a few significant  line of sight (LoS) and non-LoS (NLoS) paths between Tx and Rx, hereafter referred to as a \textit{path skeleton} \cite{sur2016beamspy}. Due to the lack of SCR knowledge, Tx/Rx in the beam-searching phase scans all directions exhaustively to find the best beamforming/combining direction. A disadvantage of this method is that the overhead increases almost linearly  with the number of beam directions. The second property of SCR is the strong correlation of the values in proximity locations as the Rx experiences almost the same scattering environment \cite{sun2018propagation,sur2016beamspy,8057188}. This is also known as the robustness of the channel second-order statistics to the small mobility \cite{shokri2015millimeter}.

Spatial correlation of mmWave SCRs have received a considerable attention  in the recent years in order to alleviate the beamforming overhead. For example, 
Sur \textit{et al.} in \cite{sur2016beamspy} presented a model that captures spatial and blockage-invariant correlation among beams to predict beam directions when human bodies block the links. However, this model is based on static links and it cannot capture the Rx mobility and is limited to an indoor environment.

Zhou \textit{et al.} in \cite{8057188} proposed a beam steering approach in an office environment that leverages the correlation of the mmWave spatial channel in near locations to predict and re-establish the blocked links. Although the results are promising in terms of throughput gain, the computational complexity of the proposed approach makes it impractical in real-world scenarios with cheap mmWave mobile devices. Moreover, the re-execution of the beam-searching is based on a constant Euclidean distance, which is very heuristic and may vary considerably in different user mobility and networking scenarios. Indeed, as we show in this paper, having a small Euclidean distance does not necessarily imply a small/negligible change in the SCR.

The existing approaches are either not suited to an outdoor mobile network or very heuristic, hindering a proper control of the tradeoff among the instantaneous rate and beamforming update overheads in various user mobility scenarios. 

In this paper, we address the problem of the beamforming  overhead and investigate an efficient beamforming algorithm in an outdoor environment. Inspired by the path skeleton approach \cite{sur2016beamspy}, we assume that every Tx maintains a database including path skeletons of different locations in its coverage areas. First, we consider a negligible overhead to construct and update this database and develop an algorithm to track the correlation between path skeletons in different locations through a specific trajectory. We show that our algorithm requires the minimum number of coarse beam-searching in comparison to the existing methods \cite{6171799} while keeping the throughput and achieved rate in a near-optimal level. \mbox{Moreover}, we show that our approach is also efficient in terms of energy consumptions, computational and signaling complexity. 
We then characterize the cost of database maintenance and show that it is inversely proportional to the number of users. It suggests that as the number of users increases, the overhead for the database update decreases drastically and therefore we can optimize the aforementioned tradeoff at almost no cost.

The rest of this paper is organized as follows. We introduce the system model in Section \ref{SM}. We describe our proposed method to track  the correlation between path skeletons in Section \ref{AL}. We also  present the ray tracing simulation results of running our algorithm in a real urban environment in Section \ref{AL}. We consider the overhead of building and updating the database in Section \ref{DB}. The summary of our findings is presented in Section \ref{con}.
\newcommand{\Tx}{\mathrm{Tx}}
\newcommand{\Rx}{\mathrm{Rx}}
\newcommand{\Base}{\mathrm{Base}}
\newcommand{\maximize}{\mathrm{maximize}}
\newcommand{\SNR}{\mathrm{SNR}}
\newcommand{\Rate}{\mathrm{Rate}}
\newcommand{\PS}{\mathrm{PS}}
\newcommand{\argmax}{\mathrm{argmax}}
\def\TT{\mathsf{T}}
\def\HH{\mathsf{H}}
\newcommand{\Pilot}{\mathrm{Pilot}}

\textit{Notations:} Bold upper-case $\mathbf{X}$, bold lower-case  $\mathbf{x}$ and normal font $x$ denote matrices, vectors and scalars, respectively. For any vector $\mathbf{x}$ (or matrix $\mathbf{X}$),  $\|\mathbf{x}\|_{2}$, ${\mathbf{x}}^\TT$ and ${\mathbf{x}}^\HH$ are its $l_{2}$-norm, transpose, and conjugate transpose, respectively. For any integer L, we define set
$[L]=\{1,2,...,L\}$. $\mathbf{I}_{L}$ denotes an $L\times L$ identity matrix.
\section{System Model}\label{SM}
\begin{figure}[t]
	\centering
	\includegraphics[width=0.32\textwidth]{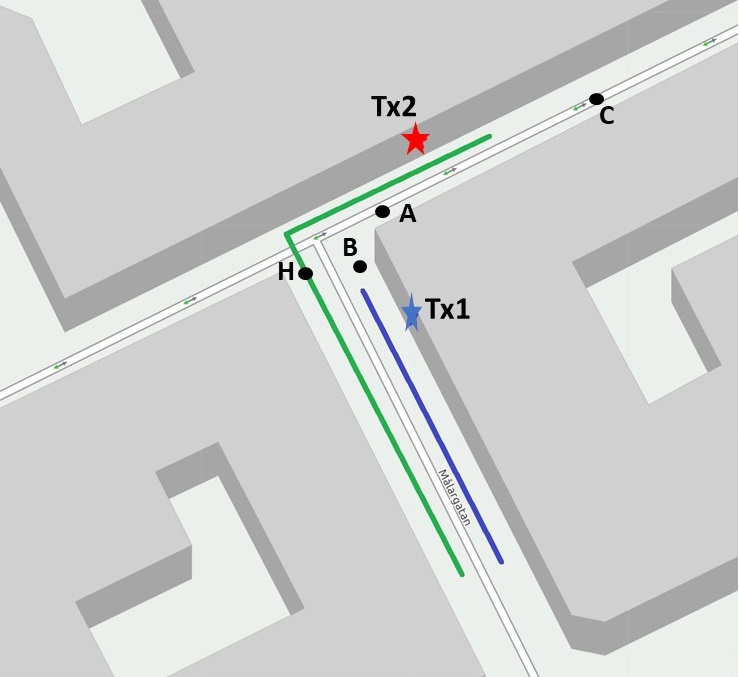}
	\caption{Simulation area in the central part of Stockholm city. The blue star shows the location of the Tx1 and the red star shows the location of the Tx2. Blue and green lines illustrate the first and the second trajectory respectively. Point H shows the handover location.}
	\label{fig:fig1}
\end{figure}

We consider the downlink of a mmWave network with multiple base stations (BSs) and user equipment (UEs). Without a loss of generality and to keep the notations simple, we focus on a specific UE and its serving BS. We consider BS as the Tx and UE as the Rx. We assume that the Tx location is constant but the Rx is mobile and moves through a specific trajectory. Blue and green lines in Fig.~\ref{fig:fig1} show different Rx trajectories.

We assume that Tx and Rx utilize uniform linear arrays (ULA) with $N_{\Tx}$ and $N_{\Rx}$ antennas, respectively. The separation between both transmit and receive antenna elements is $\lambda/2$, where $\lambda$ is the wavelength. We consider one radio-frequency chain both in the Tx and in the Rx. The channel follows a narrow band cluster model with $L$ paths, and block fading, where the small scale fading is constant over a coherent interval (CI). The channel matrix $\mathbf{H}\in\mathbb{C}^{N_{\Rx}\times N_{\Tx}}$ in one CI can be expressed as \cite{akdeniz2014millimeter}:
\begin{equation}
\mathbf{H}_{(x_{i},y_{i})}=\sqrt{\frac{N_{\Tx}N_{\Rx}}{L}}\sum_{\ell=1}^{L} \bar{g}_{\ell i} e^{j2\pi f_{d_{\ell i}}} \mathbf{a}_{\Rx} (\theta_{\ell i}) \mathbf{a}_{\Tx}^{\HH}(\phi_{\ell i}),
\label{H}
\end{equation}
where $\bar{g_{\ell i}}\sim\mathcal{CN}(0,10^{-0.1PL_{i}})$ is the complex gain of the $\ell$-th path that includes path loss and small-scale fading. $PL_{i}$ is the omnidirectional path loss that is a function of the distance between the Tx and a specific Rx in location $(x_{i},y_{i})$ where $i$ is the location index. $f_{d_{\ell i}}$ is the Doppler shift of the $\ell$-th path that is characterized by the   the direction of received paths relative to the motion of the Rx in location index $i$ (see \cite{akdeniz2014millimeter,heath2016overview} for more details). 
$\mathbf{a}_{\Tx}\in\mathbb{C}^{N_{\Tx}}$ and $\mathbf{a}_{\Rx}\in\mathbb{C}^{N_{\Rx}}$ are unit-norm array response vectors at the Tx and the Rx, respectively.  

Since we use ULA, $\mathbf{a}_{\Tx}(\phi_{\ell i})$ is \cite{kutty2016beamforming}
\begin{equation}
\mathbf{a}_{\Tx}(\phi_{\ell i})=\dfrac{1}{\sqrt{N_{\Tx}}}[1,e^{j\pi \sin(\phi_{\ell i})},...,e^{j\pi(N_{\Tx}-1) \sin(\phi_{\ell i})}]^{\HH},
\end{equation}
where $\phi_{\ell i}\in [0,2\pi)$ and $\theta_{\ell i}\in [0,2\pi)$ are angle of departure (AoD) and angle of arrival (AoA) of the $\ell$-th path in the location index $i$, respectively.
The array response vector at the Rx, $\mathbf{a}_{Rx}(\theta_{\ell i})$, can be expressed in a similar fashion by replacing $N_{\Tx}$ and $\phi_{\ell i}$ by $N_{\Rx}$ and $\theta_{\ell i}$, respectively.  
For the sake of notation simplicity, we drop subscript $(x_{i},y_{i})$ from $\mathbf{H}$ when it is clear from the context.

With the unit-norm beamforming vector $\mathbf{f}\in\mathbb{C}^{N_{Tx}}$ in the Tx and the unit-norm combining vector $\mathbf{w}\in\mathbb{C}^{N_{Rx}}$ in the Rx, the received signal of path $\ell$ in one CI is
\begin{equation}
y_{\ell}=\mathbf{w}_{\ell} \mathbf{H} \mathbf{f}^{\HH}_{\ell} \mathbf{s}+\mathbf{w}^\HH_{\ell} \mathbf{n}_{\ell},
\end{equation}
where $\mathbf{s}$ is the transmitted symbol and  $\mathbf{n}_{\ell} \sim \mathcal{CN}(0,\sigma^2)$
is the Gaussian noise vector. For the sake of simplicity, we consider $\mathbf{s}$ as the $\ell$-th column of $\sqrt{P/L}  \mathbf{I}_{L}$ where the total transmission power $P$ is equally divided among $L$ paths. Since $\mathbf{w}_{\ell}$ is unit-norm, we have that $\mathbf{w}^H_{\ell} \mathbf{n}_{\ell} \sim \mathcal{CN}(0,\sigma^2)$.
We express the beamforming vector, $\mathbf{f}$, and the combining vector, $\mathbf{w}$, as
\begin{equation}
\begin{aligned}
\mathbf{f}(\phi, N_{\Tx})=\dfrac{1}{\sqrt{N_{\Tx}}}[1, e^{j\pi \sin(\phi)},...,e^{j\pi (N_{\Tx}-1)\sin(\phi)}]^\HH,
\end{aligned}
\label{f}
\end{equation}
\begin{equation}
\begin{aligned}
\mathbf{w}(\theta, N_{\Rx})=\dfrac{1}{\sqrt{N_{\Rx}}}
[1, e^{j\pi \sin(\theta)},...,e^{j\pi (N_{\Rx}-1)\sin(\theta)}]^\HH .
\end{aligned}
\label{w}
\end{equation}
Equations (\ref{f}) and (\ref{w}) indicate that only the phase of elements in beamforming and combining vectors can be varied. 
We only consider horizontal (2D) beamforming/combining and the extension to 3D is straightforward. Each path can be LoS or NLoS.

We define a path skeleton as a set of $L$ paths that approximate the SCR between the Tx and Rx.  The path skeleton can be obtained after running exhaustive beam-searching in all directions and applying  the so-called path skeleton construction procedure in \cite{sur2016beamspy}. The Tx and Rx then search over this skeleton, instead of an exhaustive search, to substantially reduce the cost of channel estimation and beamforming design. We assume that Tx (BS) can store the path skeletons in different locations of its coverage area in a database, which should be regularly maintained. However, this task involves two costs: the maintenance cost and the query cost. The maintenance cost is the overhead of building and updating the database. The query cost is the number of times we inform the Rx about a new  path skeleton.  
We first focus on the case where the maintenance cost is negligible  and optimize the Rx (user) rate experience with a given limited query budget. 
We then relax the negligible maintenance cost and propose a simple approach to update the database, whose cost is a decreasing function of the number of Rxs.

Formally, we define a path skeleton based on the large-scale parameters of each path $p_{\ell}=\{\theta_{\ell},\phi_{\ell},\tilde{g}_{\ell}\}$, $\ell\in[L]$ between a Tx in a fixed location and a mobile Rx in location $(x_{i},y_{i})$ as
\begin{equation}
\mathbf{\PS}_{(x_{i},y_{i})}=\sqrt{\frac{N_{\Tx}N_{\Rx}}{L}}\sum_{\ell=1}^{L} \tilde{g}_{\ell i} \mathbf{a}_{\Rx} (\theta_{\ell i}) \mathbf{a}_{\Tx}^{H}(\phi_{\ell i})
\label{PS}
\end{equation}
where $\tilde{g}_{\ell i}=\sqrt{PL_{i}}$. At this location, the large-scale parameters $\theta_{\ell i}$, $\phi_{\ell i}$ and $\tilde{g}_{\ell i}$ correspond to the AoA, AoD and the gain of the $\ell$-th path.

In this work, we omit the interference effects of other Txs and leave it as the future work.

\section{Proposed Algorithm with Negligible Database Maintenance Cost}
\label{AL}
In this section, we consider the low overhead $\mathbf{\PS}$ database maintenance   scenario. This assumption essentially means all the $\mathbf{\PS}$s are available a priori at every Tx for all locations inside its coverage areas. 
Consequently, we only focus on the query cost and develop optimization problems to maximize the Rx instantaneous rate given a fixed query budget. In summary, our solution approach  measures the change in the $\mathbf{\PS}$s as the Rx moves and initiates a query automatically only when a substantial change in $\mathbf{\PS}$s, and consequently in the optimal beamforming vectors, is observed. We will consider the database maintenance cost in Section \ref{DB}.


Notice that our proposed algorithm is based on the downlink transmission; however, it can be applied to the uplink  case by replacing the roles of the beamformer ($\mathbf{f}$) and combiner ($\mathbf{w}$) and modifying the channel matrix $\mathbf{H}$.  

\subsection{Beamforming Design}
In the pilot transmission phase, the Rx asks the database for the $\mathbf{\PS}$ of its current location $(x_{i},y_{i})$. Then, a sequence of pilot signals, $\Pilot_{\ell i }=(\frac{P}{L}, \mathbf{f}(\phi_{\ell i}), \mathbf{w}(\theta_{\ell i})), \ell\in [L]$, are sent along $L$ paths in the skeleton, whose cardinality is much smaller than the total number of paths between the Tx and Rx, to find  the existing non-blocked paths. More specifically, Rx measures signal strength in all $p_{\ell} \in \PS_{(x_{i},y_{i})}$ and constructs $\mathbf{H}$ \mbox{from (\ref{H})}. If all paths $p_{\ell} \in \PS_{(x_{i},y_{i})}$ are blocked or weakened due to the presence of some potential obstacles or the dynamics of the environment, the path skeleton finder procedure is called to refresh this entry ($\PS_{(x_{i},y_{i})}$) of the database.

In the data transmission phase, the beamforming vector $\mathbf{f}$ and combing vector $\mathbf{w}$ are designed to maximize the link budget, namely:

\begin{subequations}
	\begin{equation}
	\begin{aligned}
	& \underset{\mathbf{f}, \mathbf{w}}{\text{maximize}}
	&& |\mathbf{w}^\HH\mathbf{H}\mathbf{f}|^2 
	\end{aligned}
	\end{equation}
	\begin{equation}
	\begin{aligned}
	& \text{subject to}
	&& \mathbf{f} \in \mathcal{F}, 
	\end{aligned}
	\end{equation}
	\begin{equation}
	\begin{aligned}
	&&&&&&&&&&\mathbf{w} \in \mathcal{W},
	\end{aligned}
	\end{equation}
	\label{codebook}
\end{subequations}
\kern-.5em where $\mathcal{F}$ and $\mathcal{W}$ are predefined beamforming and combing codebooks, respectively. Vectors $\mathbf{f}$ and $\mathbf{w}$ follow the functions given by (\ref{f}) and (\ref{w}). When there is no restriction on using any phase shift $\phi$ at the Tx and $\theta$ at the Rx and the number of antenna elements ($N_{\Tx}$ and $N_{\Rx}$) grows large, the optimal solution of (\ref{codebook}) is identical to the antenna response toward the strongest path. That is, $\mathbf{f}^\star=\mathbf{a}_{\Tx}(\phi_{\ell})$ and $\mathbf{w}^\star=\mathbf{a}_{\Rx}(\theta_{\ell})$, where
$\ell=\mathrm{argmax}_{\ell} \: \bar{g}_{\ell i}$.
Notice that $\bar{g}_{\ell i}$, $\ell\in[L]$ are found in the channel estimation over the $\mathbf{\PS}$. In summary, finding and applying the optimal beamforming in the asymptotic regime is very efficient: Rx should feed back to Tx the index of the path with maximum received signal strength. In this paper, we use this approach, which is asymptotically optimal in the sense of (\ref{codebook}). 

Given designed $\mathbf{f}_{\ell}$ and $\mathbf{w}_{\ell}$, the received $\SNR$ follows
\begin{equation}
\SNR=\dfrac{P|\mathbf{w}_{\ell}^\HH\mathbf{H}\mathbf{f}_{\ell}|^2}{ B  N_{0}}, 
\end{equation} 
where, $N_{0}$ is the noise spectral density and $B$ is the signal bandwidth in the data transmission phase.  
The achievable rate per second is then $\Rate=B\log(1+\SNR)$ and we can find the achievable throughput in location index $i$ by multiplying $\Rate_{i}$ by the reaming data transmission time. A faster beam-searching leads to a longer data transmission time and perhaps a higher achievable throughput. 
\subsection{Tracking Spatial Correlation}

To run the channel estimation (or equivalently beam-searching over the skeleton) on every new location, the Tx should inform the receiver about the set of paths in the skeleton. However, for most mobility models, the skeleton is almost the same over many CIs, essentially over several consecutive locations of the trajectory. To reduce unnecessary skeleton inquiry and feedback overhead, Tx continuously monitors the variations of the skeletons in different locations and reports the new one only when a significant change is detected. Formally, assume that Tx has already reported $\mathbf{\PS}_{(x_{0},y_{0})}$ as the reference skeleton. For any new location $(x_{i},y_{i})$, we define 
\begin{equation}
d(x_{i},y_{i} ; x_0,y_0) = \|\mathbf{\PS}_{(x_{i},y_{i})}-\mathbf{\PS}_{(x_{0},y_{0})}\|_{2}.
\end{equation}
Rx and Tx know $\mathbf{\PS}_{(x_{0},y_{0})}$ and $\mathbf{H}_{(x_{0},y_{0})}$. At any new location $(x_{i},y_{i})$, they use the skeleton of $(x_{0},y_{0})$ to estimate $\mathbf{\PS}_{(x_{i},y_{i})}$ and $\mathbf{H}_{(x_{i},y_{i})}$. Then, Tx translates condition $d(x_{i},y_{i} ; x_0,y_0) \leq T$ to the validity of the existing skeleton. The beamforming vectors for $(x_{i},y_{i})$ match to the strongest path in $\mathbf{H}_{(x_{i},y_{i})}$. If $d(x_{i},y_{i} ; x_0,y_0) > T$, Tx detects a significant change in the dominant paths of the channel, informs the Rx about the new skeleton, and asks for the beam-searching over the new skeleton. Then, this new skeleton is considered as the reference skeleton. This condition can be interpreted as if a set of pilot messages based on $\PS_{(x_{i},y_{i})}$ are sent to the Rx in location index $j$. Since the distance between skeletons in location index $i$ and $j$ is higher than the threshold $T$, additional terms $\mathbf{a}_{\Tx}^{H}(\phi_{\ell j})\mathbf{a}_{\Tx}^{H}(\phi_{\ell i})$  appear in (\ref{PS}) and cause worse channel estimation result.

The performance of the proposed algorithm is dependent on the mobility model, network topology, and the decision threshold $T$. Among them we can only control $T$. Higher thresholds leads to fewer overheads of the feedback channel to send the new skeleton but also the adoption of suboptimal beamforming solutions: $\mathbf{f}$ and $\mathbf{w}$ are designed based on the skeleton in reference location ($\PS_{(x_{0},y_{0})}$) not necessarily the skeleton at the current location $(x_{i},y_{i})$. Lower $T$ improves the rate performance at the expense of higher feedback overhead. To properly set this hyper-parameter of the algorithm, we run the proposed algorithm on a dataset of various trajectories and find an optimal $T$ from the following optimization problem: 
\begin{subequations}
	\begin{equation}
	\begin{aligned}
	& \underset{T>0}{\text{${T^\star}$ = $\mathrm{argmax}$}}
	\sum_{i \in [M]}\Rate_{i}(T) 
	\end{aligned}
	\end{equation}
	\begin{equation}
	\begin{aligned}
	& \text{subject to}
	&& U<U_{max} .
	\end{aligned}
	\end{equation}
	\label{T}
\end{subequations}
\kern-.5em where $U_{\max}$ is a maximum tolerable number of skeleton queries (database query budget) and $M$ is the length of the trajectory. The optimization problem (\ref{T}) includes a \mbox{one-dimensional search} over $T>0$ and can be solved numerically. Examples of the trajectories in the training dataset are shown in Fig.~\ref{fig:fig1}. Once $T$ is obtained, this can be applied to users with similar mobility patterns.  

Reference \cite{8057188} considers $d(x_{i},y_{i};x_0,y_0)=\sqrt{(x_{i}-x_{0})^2+(y_{i}-y_{0}^2)}$, Euclidean distance, to assess the $\mathbf{\PS}$ correlation in different locations. This metric is both mobility-agnostic and topology-agnostic. To illustrate, consider the following example. Let Tx2 serves an Rx that is moving from location A to location B, see Fig.~\ref{fig:fig1}. Due to the small distance between location A and B (about 5~m), the Euclidean approach declares that the  spatial channels in location A and B are highly correlated. However, the presence of a building obstacle between two locations completely changes the AoAs and AoDs and consequently  the spatial channels. Thus, skeletons in location A and location B are not correlated. However, our proposed metric considers the actual distance between spatial channels (in norm-2 sense) and tracks the validity of the existing paths between the Tx and the Rx. Now, consider the Rx  moves from location A to location C that is about 15~m far away. Although based on the Euclidean distance approach the spatial channel correlation weakens due to the long distance between the two locations, the AoAs and AoDs  in the two locations are quite similar so SCRs and equivalently $\mathbf{\PS}$ should be indeed correlated. Hence, the norm-2 distance of channels as the metric to track spatial correlation in different locations can be much more accurate than a simple Euclidean distance approach.

\begin{algorithm}[tp] 
	\caption{Tracking spatial correlation}\label{Alg1}
	\textbf{Inputs:}  A $\mathrm{Trajectory}=\{(x_{1},y_{1}),...,(x_{M},y_{M})\}$ of M coordinates, $T$ and $B$. 
	\begin{algorithmic}[1]
		\State{Initialization: Set $(x_{0},y_{0})\leftarrow (x_{1},y_{1}),  
			\PS_{0}\leftarrow$ skeleton at ${(x_{1},y_{1})}$ and $U=0$ }
		\For {$i=1,...,M$}
		\State {Send pilots over $\PS_{(x_{0},y_{0})}$ and observe $\{\bar{g}_\ell\}_{\ell \in [L]}$}
		\If{$d(x_{i},y_{i};x_{0},y_{0})<T$}
		\State {$\mathbf{f}^\star = \mathbf{a}_{\Tx}(\phi_{\ell}), \ell = \mathrm{argmax}_\ell \:  \bar{g}_\ell$} 	
		\State {$\mathbf{w}^\star=\mathbf{a}_{\Rx}(\theta_{\ell}), \ell = \mathrm{argmax}_\ell \:  \bar{g}_\ell$} 
		
		\Else
		
		\State { // Update reference point}
		
		\State {$(x_{0},y_{0}) \leftarrow (x_{i},y_{i})  $}
		\State{$ \PS_{(x_{0},y_{0})}\leftarrow$  ${\PS_{(x_{i},y_{i})}}$ and inform user }
		\State{Send pilots over $ \PS_{(x_{0},y_{0})}$ }
		\State {$\mathbf{f}^\star = \mathbf{a}_{\Tx}(\phi_{\ell}), \ell = \mathrm{argmax}_\ell \:   \bar{g}_\ell$} 	
		\State{$\mathbf{w}^\star=\mathbf{a}_{\Rx}(\theta_{\ell}),  \ell = \mathrm{argmax}_\ell \:   \bar{g}_\ell$} 
		\State {$U\leftarrow U+1 $}
		
		\EndIf
		
		\State{$\Rate_{i}=B\log(1+\SNR_{i})$}
		\EndFor
		\\
		\textbf{Outputs:} $\Rate_{i}(T)$ and $U$
	\end{algorithmic}
\end{algorithm}

\subsection{Efficiency of the Proposed Algorithm}
Efficiency can be defined based on four parameters: computational complexity, signaling complexity, throughput efficiency, and energy consumption.
In the context of \mbox{beamforming for} mmWave networks, an efficient  algorithm should keep the throughput and energy consumption at an optimal level with manageable computational and signaling complexities.

Given the presence of an updated skeleton database, \mbox{Algorithm \ref{Alg1}} adds negligible numbers of computations to the system. However, this computation uses historical data about the environment to decrease the number of running coarse \mbox{beam-searchings in} all directions and they trigger if the dynamics of the environment change the skeletons in the database. Moreover, based on our proposed beamforming design, in order to find the existing paths between Tx and Rx in reference locations, our algorithm searches only over the skeleton instead of all the space, leading to substantially less signaling complexity.

In terms of throughput, our algorithm checks the distance between spatial channels in different locations and  updates the beamforming and combining vectors if they have been designed based on an obsolete channel, thereby guaranteeing a close-to-optimal performance of the channel estimation and beamforming design. Moreover, due to a faster beamforming design, we expect a longer data transmission time and therefore a gain in the achievable throughput.

In terms of energy efficiency, our approach keeps the number of pilots at a minimal level by searching only over the skeleton and reducing the database query frequency. Thus our approach is energy efficient, as we observe in the numerical results.  
\begin{figure}[!t]
	\begin{subfigure}[t]{0.5\textwidth}
		\includegraphics[width=0.97\textwidth]{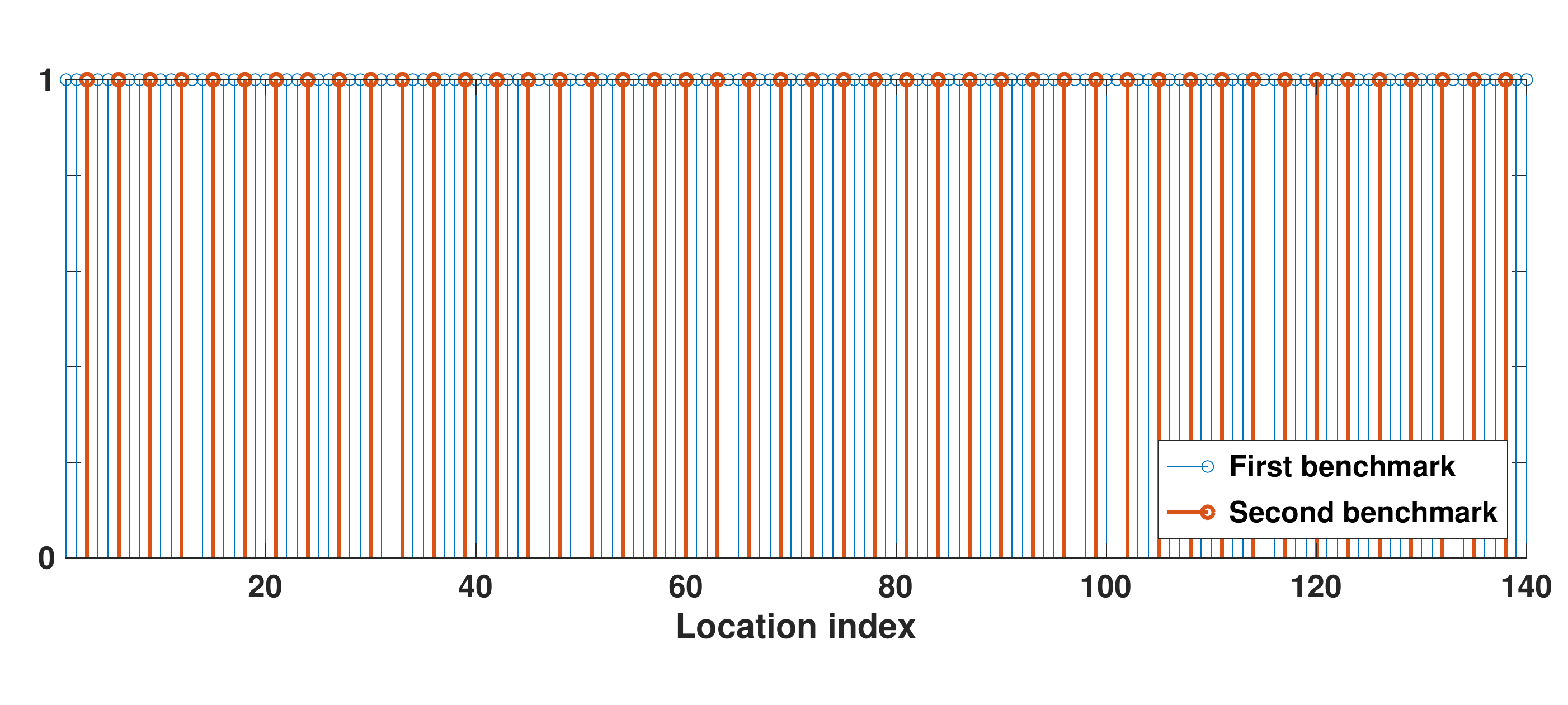}
		\caption{}
		\label{fig2:(a)}
	\end{subfigure}	
	\begin{subfigure}[t]{0.5\textwidth}
		\includegraphics[width=0.97\textwidth]{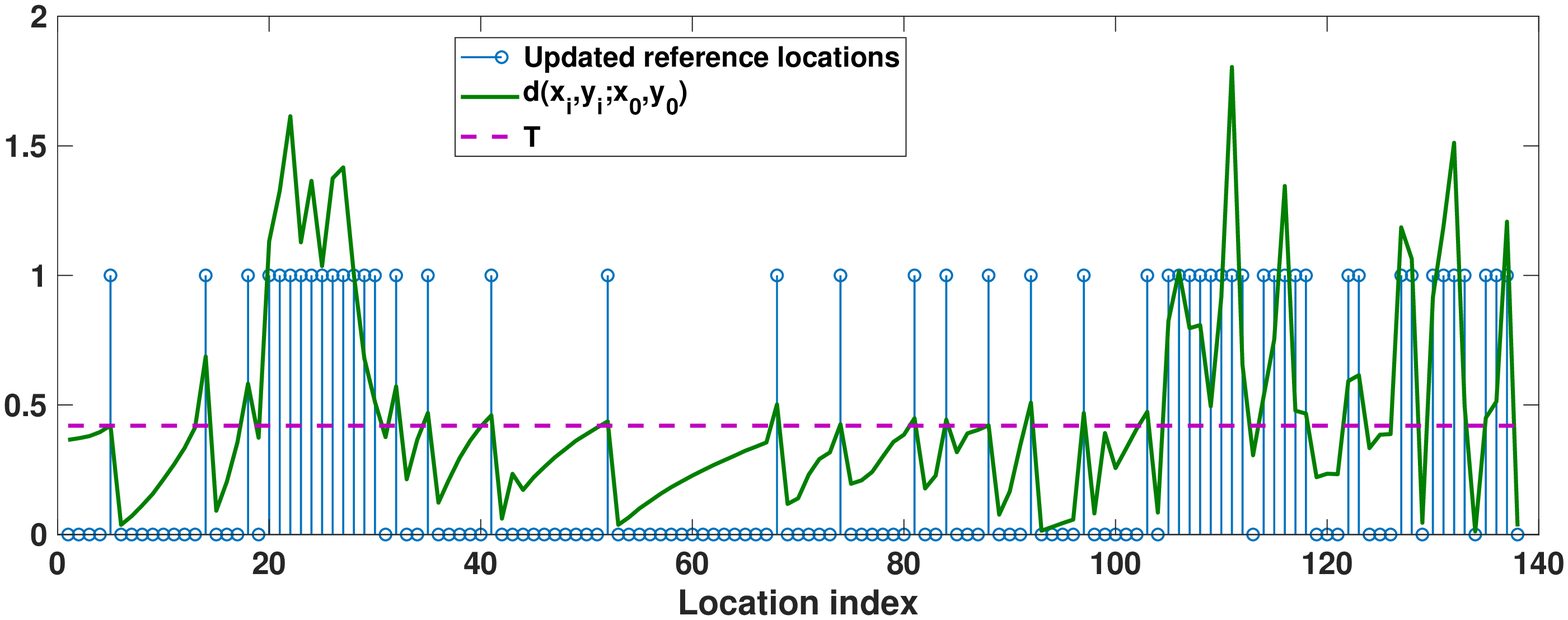}
		\caption{}
		\label{fig2:(b)}
	\end{subfigure}
	\caption{The index of the updated reference locations by running (\subref{fig2:(a)}) benchmarks, and (\subref{fig2:(b)}) Algorithm \mbox{\ref{Alg1}}.}
	\label{fig2:2}
\end{figure}
\subsection{Numerical Results}
We numerically evaluate the performance of our proposed approach in an urban environment using a ray tracing tool \cite{simic2017demo}. From the ray tracing output, we can obtain the existing  paths  between the Tx and the Rx in a specific location. To ensure high angular resolutions, we measure the AoAs and the AoDs with step sizes of 0.1 degrees. We extract the 200 m$\times$200 m real building map of a central part of Stockholm city and use it as the input data for the ray tracing simulation. Fig.~\ref{fig:fig1} shows the simulation area. We randomly assign glass or brick materials to the buildings. The general simulation parameters are listed in Table \ref{table1}. We consider two different Rx trajectories. We assume a normal pedestrian walk with speed 5 km/h.

\begin{table}[b]
	\centering
	\caption{Simulation parameters.}
	\begin{tabular}{ccc}\hline
		BS transmit power & 30 dBm  \\
		Path loss exponent & 3  \\
		
		Operating frequency & 28 GHz \\
		Signal bandwidth in data transmission phase & 500 MHz \\
		Thermal noise power & -174 dBm/Hz \\
		Rx height & 1.5 m \\
		Tx height & 6 m	\\
		Number of Tx antennas & 8 \\
		Number of Rx antennas & 8 \\
		Brick penetration loss \cite{zhao201328} & 28.3 dB \\
		Glass penetration loss \cite{zhao201328} & 3.9 dB    
		\\\hline 
	\end{tabular}
	\label{table1}
\end{table}
\begin{figure}[!t]
	\begin{subfigure}[t]{0.5\textwidth}
		\includegraphics[width=0.97\textwidth]{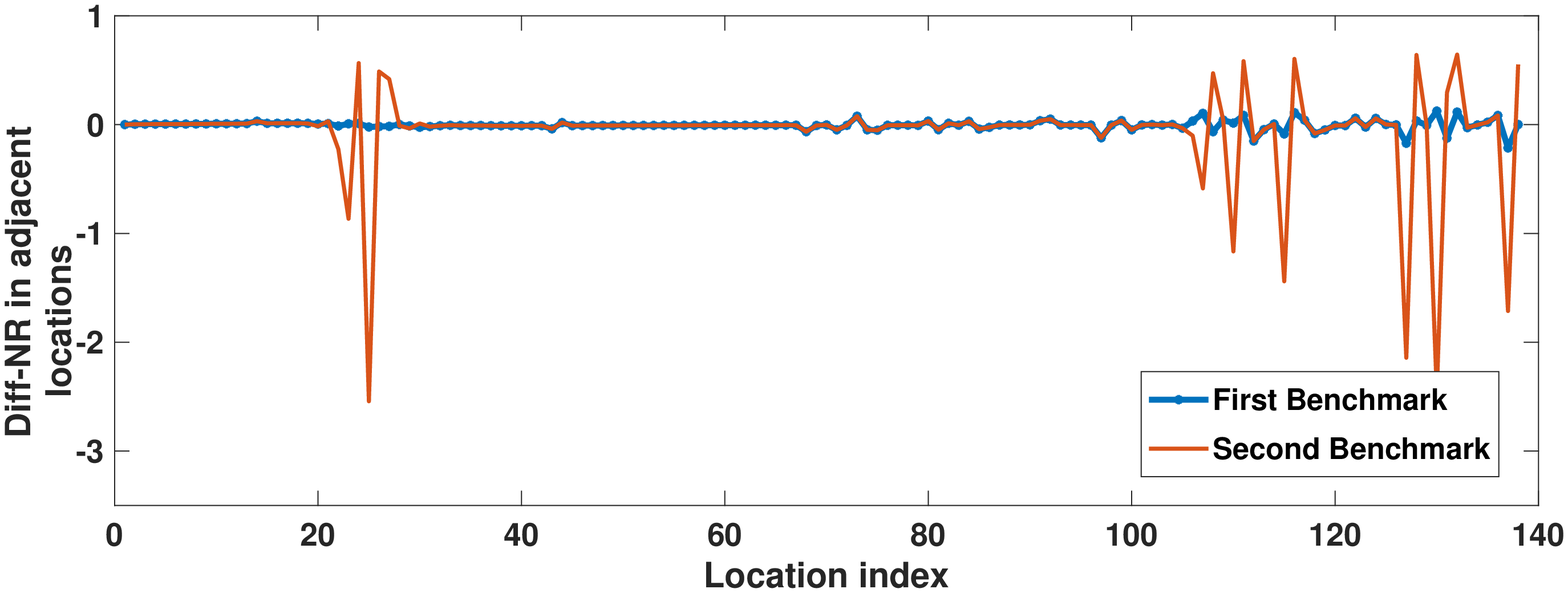}
		\caption{}
		\label{fig3:a}
	\end{subfigure}	
	\begin{subfigure}[t]{0.5\textwidth}
		\includegraphics[width=0.97\textwidth]{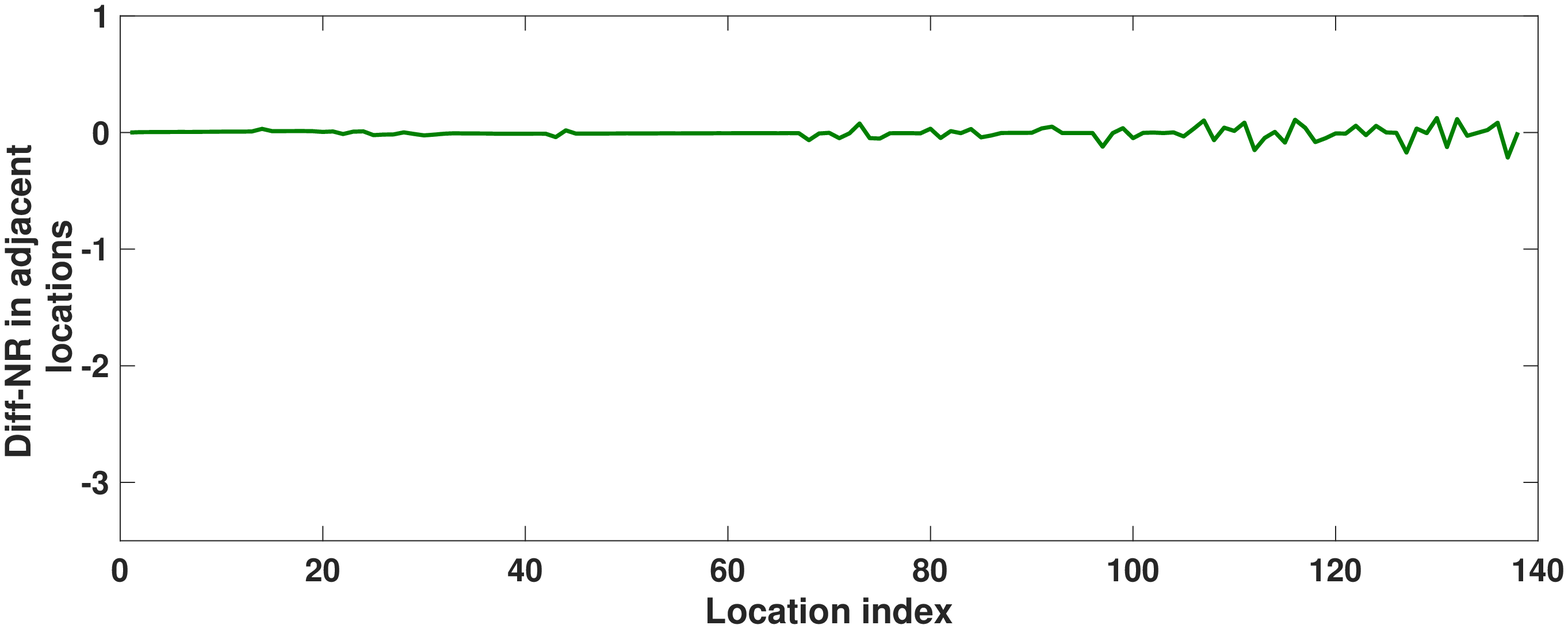}
		\caption{}
		\label{fig3:b}
	\end{subfigure}
	
	\caption{Diff-NR in adjacent locations based on (\subref{fig3:a}) benchmarks, and (\subref{fig3:b}) \mbox{Algorithm \ref{Alg1}}.}
	\label{fig3:3}
\end{figure}
\subsubsection{First Scenario}
In this scenario, we consider one Tx on the wall of a building with a height of 6 m and blue Rx trajectory of Fig.~\ref{fig:fig1}. Rx is served by Tx1 along this trajectory. The trajectory length is 140 m, including 140 equidistant locations (once every meter), the Euclidean  distance between Tx and Rx varies from 24 m to 119 m. In the \textit{first benchmark}, we simulate a baseline where Tx extracts the $\mathbf{\PS}$s from the database in all locations so the database query budget is equal to the length of the trajectory.
In the \textit{second benchmark}, we  simulate the approach of \cite{8057188} where the beamforming and combining directions updates based on a fixed Euclidean distance as shown in Fig.~2(a).
The distance between two consecutive updates is 3 m, which is chosen to keep the same total number of updates as our approach (50 updates budget). In Fig.~2(b), the green curve shows distance between $\PS_{(x_{i},y_{i})}$ in the location index $i$ and the reference skeleton, $\PS_{(x_{0},y_{0})}$. Based on Algorithm \ref{Alg1}, the reference location is updated once the green curve is higher than the threshold $T$, set to be 0.42 for our environment as the optimal solution of (\ref{T}) with a query budget of $U_{\max}$ = 50. We have marked the update points by
stems. The irregular inter-stem distance is due to the mobility of the user and similarity of the channels for some intervals.

We define difference normalized rate (Diff-NR) in two adjacent locations $(x_{i},y_{i})$ and  $(x_{i+1},y_{i+1})$ as 
\begin{equation}
\text{Diff-NR} = \frac{\Rate_{i+1} - \Rate_i}{\Rate_i}.
\end{equation}
In the first benchmark, clearly, Rx achieved rates in each location is maximum because of using optimal beamforming and combining directions in all locations but with the query cost of 140. In the second benchmark,
As shown in Fig.~3(a), the user experiences some high fluctuations in the instantaneous rate, which are due to the fact that the channel correlation is sometimes weak before 3 m distance and we need to update the beamforming vectors in the meanwhile. Such fluctuations may prohibit the support of a reliable connection to the user. 
Fig.~3(b) shows Diff-NR of running Algorithm \ref{Alg1}, indicating a very similar pattern as the first benchmark. In other words, Algorithm \ref{Alg1} with up to 50 queries to the skeleton database can perform as good as the first benchmark, where we have optimized the beamforming in all 140 locations of the trajectory.

\begin{figure}[!t]
	\begin{subfigure}[t]{0.5\textwidth}
		\includegraphics[width=1\textwidth]{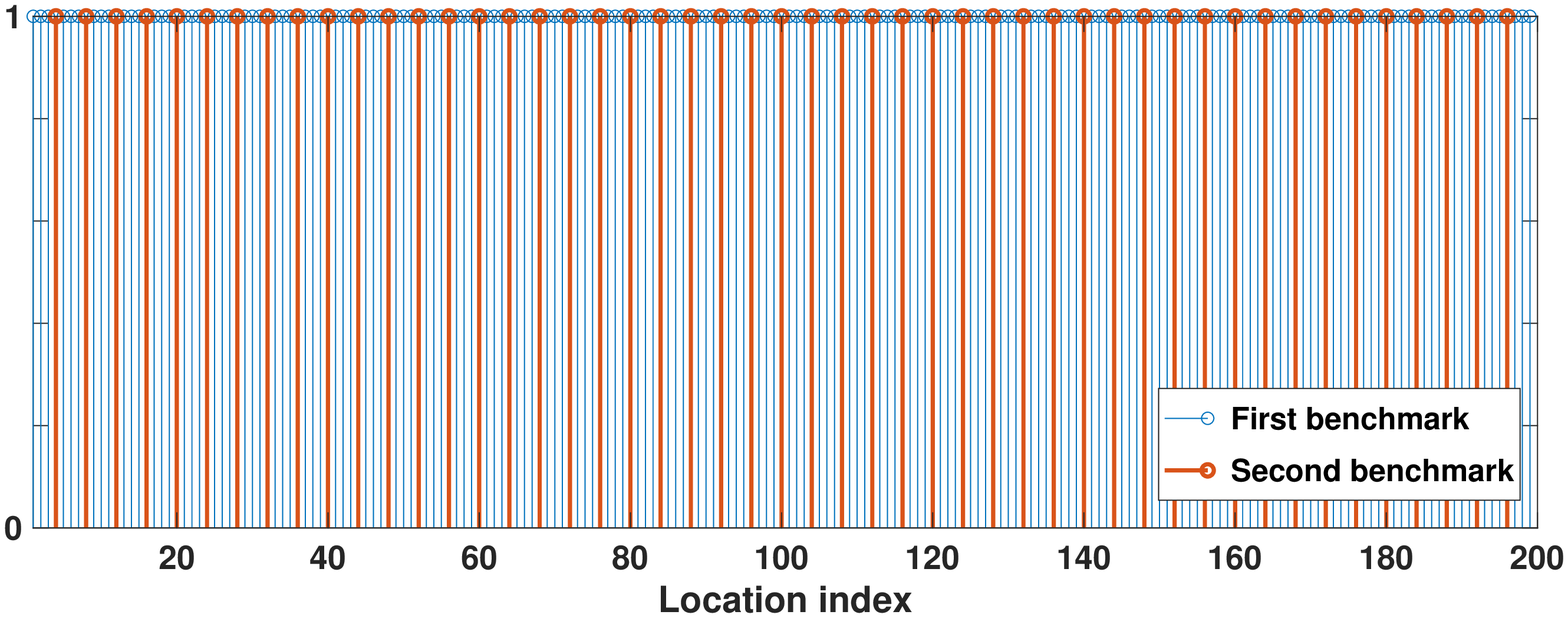}
		\caption{}
		\label{fig4:a}
	\end{subfigure}	
	\begin{subfigure}[t]{0.5\textwidth}
		\includegraphics[width=1\textwidth]{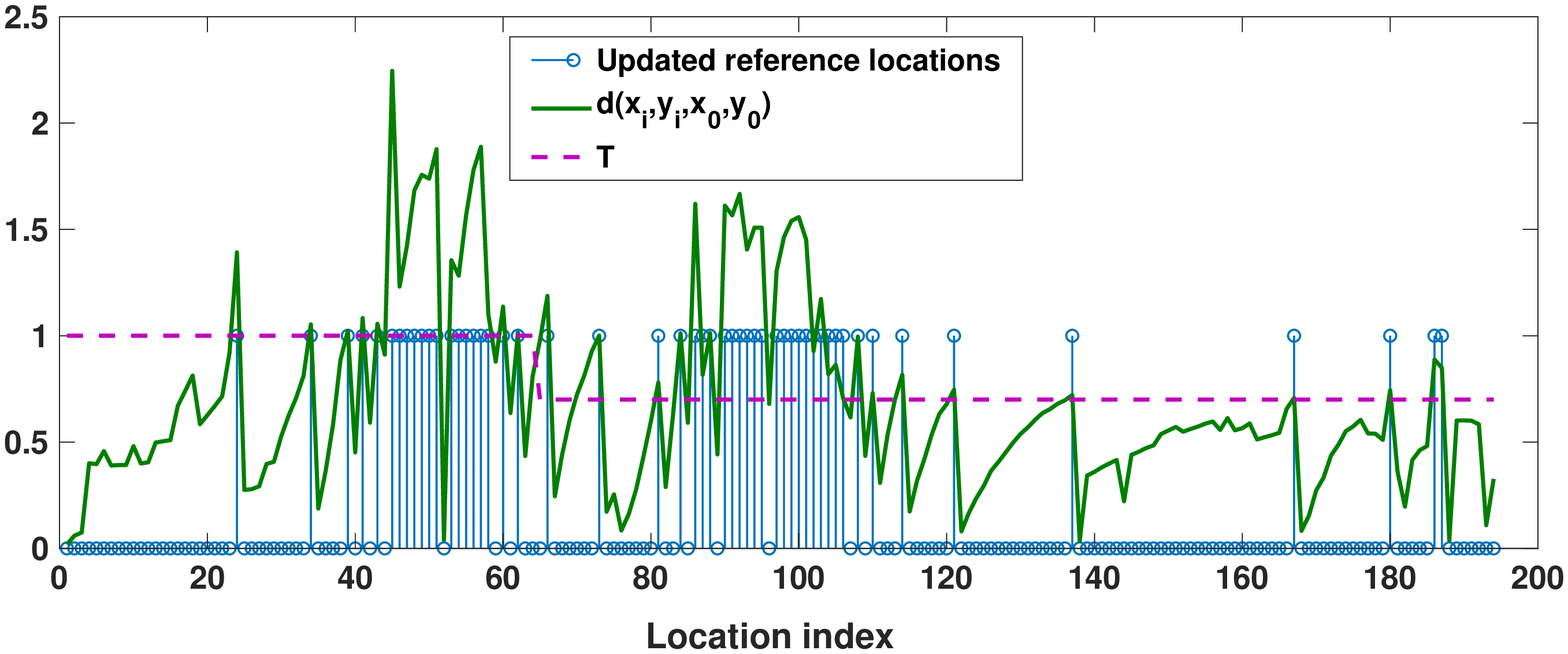}
		\caption{}
		\label{fig4:b}
	\end{subfigure}
	
	\caption{The index of the updated reference locations by running (\subref{fig4:a}) the benchmarks, and (\subref{fig4:b}) \mbox{Algorithm \ref{Alg1}}.}
\end{figure}

\subsubsection{Second Scenario}
In this scenario, there are two Txs. Green line  in Fig.~\ref{fig:fig1} shows the Rx trajectory. The length of this trajectory is 200 m and contains two parts. In the first part Tx2 serves Rx. The second part is started from point H, where Tx1 starts serving the user after a handover. Again, we consider the baseline where each Tx queries $\mathbf{\PS}$ sets from its database in all locations as the first benchmark and also consider the constant Euclidean distance query policy as a the second  benchmark as shown in Fig.~4(a). The result of runing our algorithm is presented in Fig.~4(b).  In this case, we consider two different thresholds for each Tx. Based on numerical results, we choose $T_{1}=0.7$ as the threshold of Tx1 and $T_{2}=1$ as the threshold of Tx2 that are the solutions of the optimization problem in (\ref{T}) with query budget 52. Reference locations in each Tx coverage area will be updated based on defined thresholds.

\begin{figure}[!t]
	\begin{subfigure}[t]{0.5\textwidth}
		\includegraphics[width=1\textwidth]{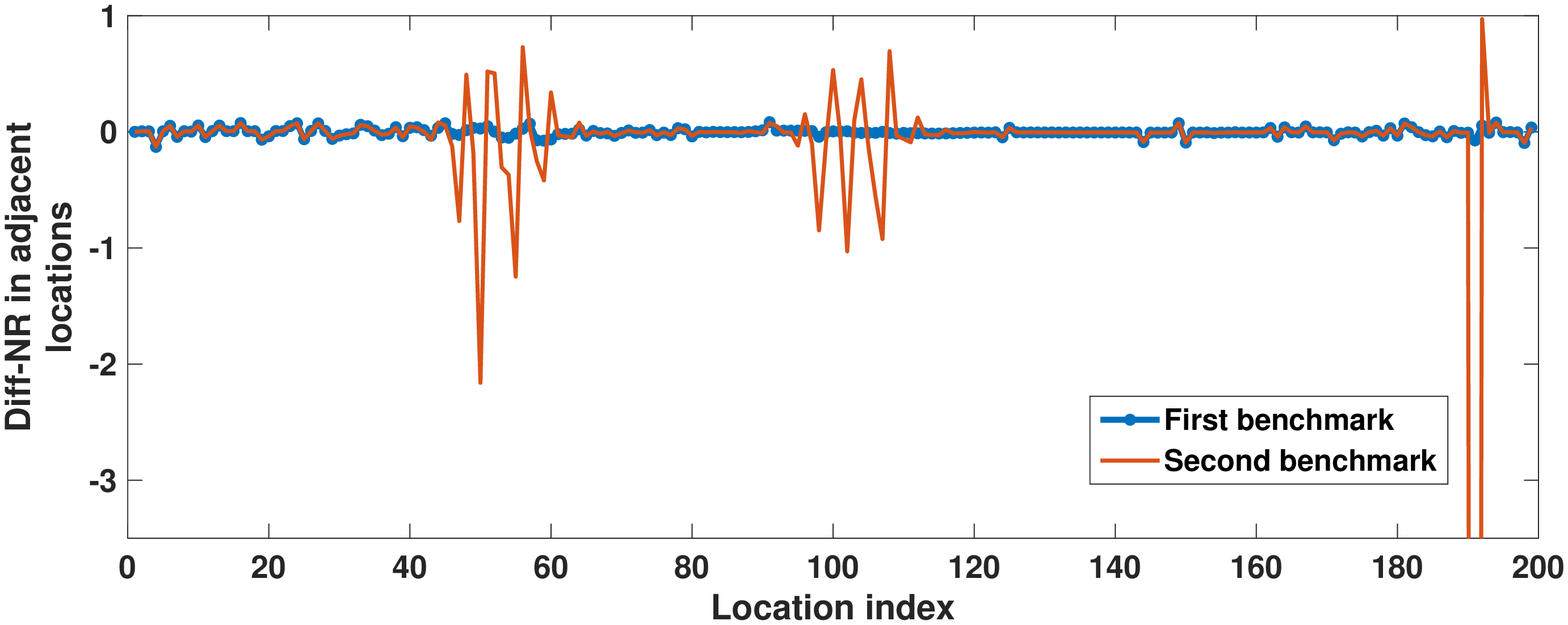}
		\caption{}
		\label{fig5:a}
	\end{subfigure}	
	\begin{subfigure}[t]{0.5\textwidth}
		\includegraphics[width=1\textwidth]{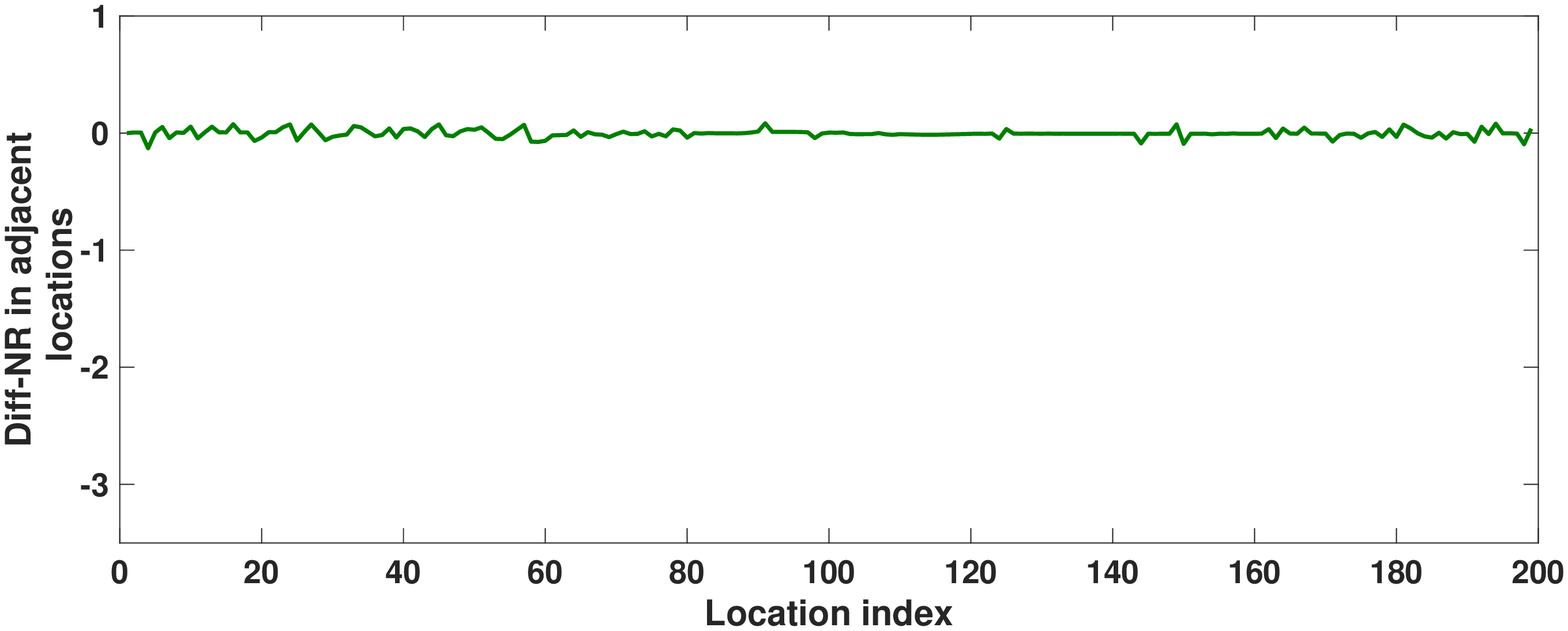}
		\caption{}
		\label{fig5:b}
	\end{subfigure}
	\caption{Diff-NR in adjacent locations based on (\subref{fig5:a}) the benchmarks, and (\subref{fig5:b}) \mbox{Algorithm \ref{Alg1}}.}
	\label{fig5:5}
\end{figure}

Fig.~\ref{fig5:5} illustrates Diff-NR in adjacent locations. It is evident that the rate fluctuation decreases after the handover point in location index of 75. In the first benchmark the query budget is 200 and the achieved rate in all locations is optimal. In the second benchmark the database query is requested every 4 m, shown in Fig.~4(a). Red curve in Fig.~5(a) shows the Diff-NR in adjacent locations that indicates multiple severe rate reductions as high as 10 times throughout the trajectory. Indeed, Rx experiences low throughput in many locations of the trajectory, further highlighting that updating beamforming and combining vectors based on a constant Euclidean distance is not an efficient solution in a dynamic environment.
Fig.~5(b) presents the results of running our proposed method, which implies that with running our algorithm and the query cost equal to 52,  Rx achieved rate is near optimal in all locations of the trajectory. 
\section{Non-negligible Database Maintenance Cost}
\label{DB}
In this section, we consider the database maintenance cost that includes the overhead of the building and updating database in a Tx. We start by describing database building and updating phases. Then, we show that the maintenance overhead is inversely proportional to the number of users. So in the outdoor environment with a large number of Rxs, the overhead of building and updating the database could be almost negligible. 
\subsection{Database Construction}
\begin{figure}[t]
	\centering
	\includegraphics[width=0.8\linewidth]{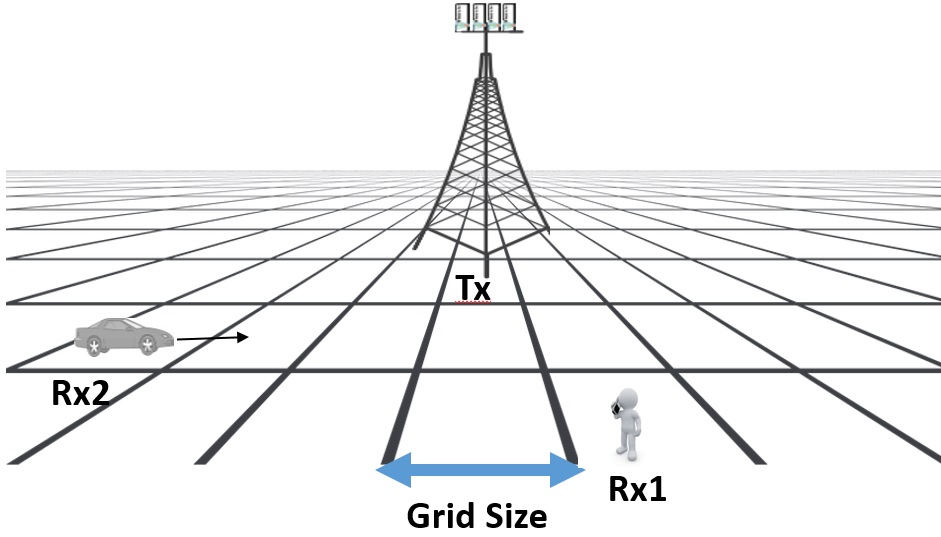}
	\caption{Tx divides its coverage area to grids with an equal size.}
	\label{fig:fig6}
\end{figure}

Initially, Tx divides its coverage area to smaller sections (grids) with pre-defined sizes as shown in Fig.~\ref{fig:fig6}. Tx assigns a unique ID to each grid, approximates each grid with one point, and considers one skeleton for that point. Therefore, the database has the same number of skeletons as the number of grid IDs. The grid size is equal in all sections and is chosen to balance the location resolution and the complexity of building a database. The beam-searching procedure runs once in each grid in order to build $\mathbf{\PS}$ set of grid IDs.

The database has two lists: the normal list and the watch list. The normal list consists of updated $\mathbf{\PS}$s. The watch list includes the grid IDs whose $\mathbf{\PS}$s should be updated. Once a user is located in such grids, Tx may run the path skeleton finder procedure and update the $\mathbf{\PS}$s for those grid IDs. Clearly, all IDs are in the watch list initially. The main steps of the initial build up of the database are as follows: 

\begin{enumerate}
	\item Tx sends a skeleton finder request to all users in every grid ID, say x. If Tx distinguishes two or more than two Rxs in  grid ID-x, it will choose one of them randomly.  
	
	\item The selected Rx confirms this request by sending a skeleton finder acknowledgment and the discover procedure (like the one in \cite{sur2016beamspy}) will be started afterward. Notice that an Rx may not confirm the request, for example, due to its low battery level. 
	\item The output of discovery process is the path skeleton in x, which includes all paths between Tx and the grid ID-x. Tx stores the skeleton, moves ID-x to its normal list, and activates an aging counter for this skeleton. The counters determine the age of each $\mathbf{\PS}$ in the normal list. 
\end{enumerate}
Tx repeats the above process for all grid IDs of the watch list and stores them in the normal list of the database. 

\begin{figure}[!t]
	\centering
	\includegraphics[width=1\linewidth]{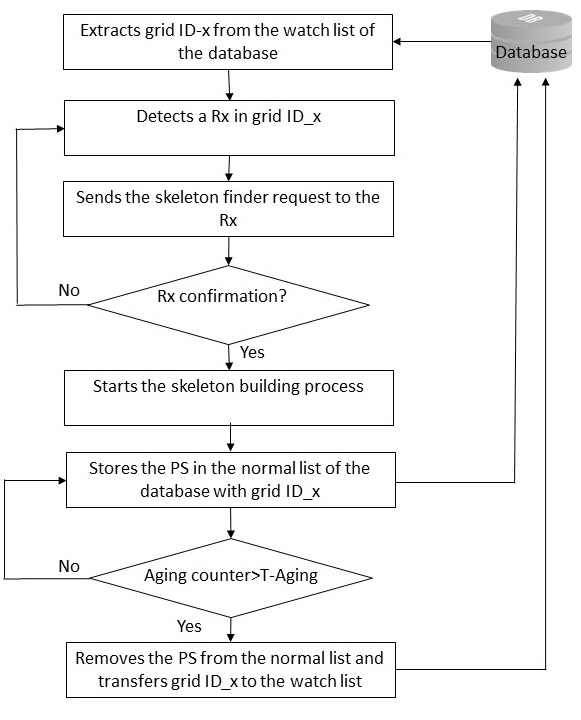}
	\caption{A flowchart of the building and the updating database. The algorithm starts by extracting the grid-IDs from the watch list of the database. When Tx detects a Rx in the grid-ID, sends the skeleton finder request to it and stores the $\mathbf{\PS}$ set in the normal list and activates the aging threshold. If the value of a aging counter exceeds the aging threshold, Tx transfers its grid ID to the watch list again. In this case, the algorithm returns to the skeleton finder loop.}
	\label{fig:fig7}
\end{figure}

In the database updating phase, Tx checks the aging counters of the $\mathbf{\PS}$s in the normal list. If an aging counter exceeds a predefined threshold (T-Aging), Tx removes the ID from the normal list and adds it to the watch list of the database. It means that the skeleton discovery process runs again for this grid ID. Fig.~\ref{fig:fig7} summarizes the 
proposed building and updating phases for a specific grid ID.

The T-Aging depends on the environment. For a highly dynamic environment like crowded streets, it should be shorter than a stationary environment, as the skeletons may change more frequently in the former situation so Tx needs to update its database in a shorter periods. 

\subsection{Database Maintenance Overhead}
In this subsection, we show that the overhead of building and updating the database is inversely proportional to the number of users in the environment. We then present numerical results of database maintenance overhead.
In the database building and updating phases the number of the times that a Tx sends the skeleton finder request to a specific Rx in a trajectory is the main metric in analyzing the database overhead. We assume that Rxs always confirm the skeleton finder requests received from the Tx.
In the dense urban environment, the number of Rxs is essentially high so the probability that one specific Rx receives several  skeleton finder requests from the Tx can be low. In other words, database maintenance overhead is distributed among all Rxs in the Tx coverage area such that the database overhead decreases as the number of the Rx increase. 

To have a better understanding of the interplay between the number of Rxs and the database maintenance overhead, we numerically evaluate the performance of the proposed method. We consider different sets of Rx in our studied environment in Fig.~\ref{fig:fig1}. We randomly assign  different trajectories to each Rx. We assume each Rx is moving through its trajectory with some random speeds and directions. For instance in Fig.~\ref{fig:fig6} we assign the normal pedestrian speed walk 5 km/h to Rx1 and average vehicle speed  30 km/h to the Rx2. Each Tx divides its coverage areas to grids with an equal size. We choose grid size equal to 2 m that is suitable for the crowded urban environment. We also consider the aging threshold equals to 2 minutes in our simulations. As  mentioned in the previous subsection, grid size and aging threshold are highly dependent on the measurement environment and need to be very small in the dense urban environment.

\begin{figure}[t!]
	\centering
	\includegraphics[width=1\linewidth]{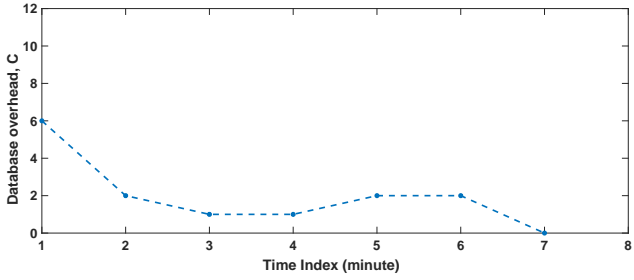}
	\caption{The value C for a Rx that is moving in a trajectory in terms of simulation time index.}
	\label{fig:fig8}
\end{figure}

We define $C$ \textit{as the database overhead in a specific time duration}. In other words, $C$ is the number of the skeleton finder requests that the Tx sends to a specific Rx in its coverage area. First, we consider 100 Rxs that are moving through different trajectories with different speeds in our studied environment. All Rxs are located in different grids in Tx1's coverage area, see Fig.~\ref{fig:fig1}. First, we run the database building and updating phases. We consider Rx1, a pedestrian, that is moving  through the trajectory that is shown with blue line in Fig.~\ref{fig:fig1} and compute $C$ during the simulation time, 7 minutes. Fig.~\ref{fig:fig8} shows the values of $C$ in different time indexes. In this case, the average amount of $C$ is 2, which means Rx1 receives only 2 skeleton finder requests on average through its trajectory. In other words, the database overhead are divided between 100 Rxs in environment so mean values of $C$ for each Rx is essentially low.  

\begin{figure}[!t]
	\centering
	\includegraphics[width=1\linewidth]{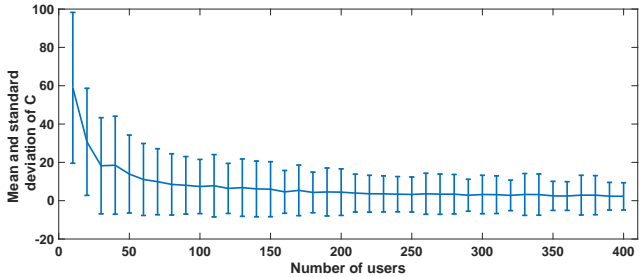}
	\caption{Average and standard deviation of $C$ for different number of Rxs/users.}
	\label{fig:fig9}
\end{figure}

Now we sweep the size of Rx sets and repeat the previous simulation for each set. We plot the average amount of $C$ for each Rx set in Fig.~\ref{fig:fig9}.  We also plot the standard deviation of values $C$ for each Rx set in Fig.~\ref{fig:fig9} that expresses how much different values of C, differ from the mean value for a specific Rx set. From Fig.~\ref{fig:fig9}, it is evident that    
as the number of Rxs increase, the mean value of $C$ and also standard deviation decreases, implying that the database overhead is almost negligible in crowded environments.

\section{Conclusion}
\label{con}
In this work, we proposed an efficient algorithm that leverages the correlation of path skeletons in order to decrease the number of running the coarse beam-searching methods in a mobile environment. We assume that each Tx is aware a priori of the path skeleton sets in its coverage area and show that the overhead of this assumption is almost negligible in dense urban environments.  
The simulation results highlight the efficiency of the proposed method and show that without significant reduction in the achieved rate, and almost the same throughput, our method can decrease the number of running beam-searching. Moreover, the results confirm that re-execution beam-searching based on a constant distance may cause low throughput while our proposed method sustains the performance in the Rx trajectory.

\bibliographystyle{IEEEtran}
\bibliography{IEEEabrv,ref_bib}

\end{document}